\documentstyle[12pt]{article}
\newcommand{\be}{\begin{equation}}
\newcommand{\ee}{\end{equation}}
\def\({\left(}
\def\){\right)}

\newcommand{\bra}[1]{\mbox{$\langle{#1}\,|$}}
\newcommand{\ket}[1]{\mbox{$|\,{#1}\rangle$}}
\newcommand{\reel}{{\rm I\kern -.2em \rm R}}
\newcommand{\naturel}{{\rm I\kern -.2em \rm N}}

\begin{document}

\title{Lifetimes of impurity states in crossed magnetic and electric fields}
\author{S\'ebastien Gyger and Philippe A. Martin \\
\small\it Institut de Physique Th\'eorique, Ecole Polytechnique F\'ed\'erale
de Lausanne, \\ \small\it CH-1015, Lausanne, Switzerland}
 \date{\today}
\maketitle
\begin{abstract}
We study the quantum dynamics of localized impurity states created by a point interaction
for an electron moving in two dimensions under the influence of a perpendicular magnetic field
and an in-plane weak electric field. All impurity states are unstable in presence
of the electric field. Their lifetimes are computed and shown to grow in a Gaussian way as
the electric field tends to zero.
\end{abstract}
\vspace{3mm}
PACS numbers: 03.65-w, 76.20+q

\section{Introduction}

A detailed understanding of the dynamics of electrons in two dimensions (2D)
in crossed magnetic and electric fields and in presence of impurity scattering 
plays an important role in the context of the quantum Hall effect and transport theory.

We consider an electron of charge $e$ and mass $m$ confined to the two dimensional plane
$(x,y)$ (without boundaries). An uniform magnetic field of magnitude $B$ acts perpendicular to the plane
and an electric field of magnitude $E$ acts in the $x$ direction. Moreover an impurity located at the origin
scatters the electron with a short range potential $V(x,y)$. When $E=0$, the electron 
remains localized in the course of the time, both for the classical and the quantum dynamics. 
Classically, modelling for instance the interaction with the impurity by a hard disk of radius $a$,
the electron accomplishes the usual circular cyclotronic motion outside of the disk or bounces
around the surface of the disk by a succession of segments of cyclotronic motion.
Quantum mechanically, the impurity potential, considered as a short range perturbation
of the pure magnetic problem (the Landau Hamiltonian), will preserve the essential spectrum
(i.e. the infinitely degenerated Landau levels) and create at most point spectrum
in-between these levels (and possibly below them if $V$ is attractive).

When the electric field is applied, the situation changes. In the classical case,
the center of the cyclotronic orbit acquires a constant drift of velocity $E/B$ in the 
$y$-direction: the trajectories around the disk get distorted by the acceleration 
imposed by the electric field and may eventually leave the disk. In the quantum case,
the Hamiltonian of the homogeneous system with the crossed magnetic and electric fields
has an absolutely continuous spectrum on $\reel$ and this spectrum will remain present after
the introduction of the short-range impurity potential, in view of general theorems
on the stability of the absolutely continous spectrum under such perturbations.
An interesting result has recently been obtained for the classical dynamics: for non zero but 
sufficiently small electric field, there exists a set of positive Lebesgue measure of trajectories 
that remain trapped near the disk [1]. The corresponding quantum mechanical question is whether
point spectrum can survive the switching-on of the electric field and remain embedded
into the continuous spectrum, provided that this field is sufficiently weak. 
To our knowledge there is no definite answer to this question at the moment.
The existence of an appreciable domain of the classical phase space supporting localized
trajectories may be an argument in favor of a corresponding localized quantum state,
but quantum interferences and tunneling phenomena may invalidate this anticipation.

In this note, we bring a partial contribution to this problem by considering
a point impurity acting as a $\delta$-potential. This model has been introduced by Prange [2]
in relation with the quantum Hall effect (with the electron in a finite strip).  
Our contribution consists of a proof that all
localized states created by the impurity are turned into resonances and of an exact determination of their lifetimes, 
shown to be of the order of
$\exp\(\frac{|B|}{|e|\hbar}\(\frac{\Delta}{E}\)^{2}\)$ as $E\to 0$ where
$\Delta$ is the distance of the resonance energy to the closest Landau level (for a more precise formula, see
section 4). However if
one allows for an impurity interaction which is not of potential form 
(i.e. an integral kernel in configuration space), one
can have point spectrum embedded into the continuum. We give an example with a rank one perturbation (section 5).  

\section{The Model}

We denote $H^{0}$ the $2D$ Hamiltonian of the electron in the crossed fields with potential
vector ${\bf A}=(0,Bx)$ in the Landau gauge 
\be
H^{0}=\frac{1}{2}p_{x}^{2}+\frac{1}{2}(p_{y}-x)^{2}-\mu x,\;\,\;p_{x}=-i\frac{\partial}{\partial x},\;\;\;
p_{y}=-i\frac{\partial}{\partial y}
\label{1}
\ee
Here $H^{0}$ is written in dimensionless variables by choosing $\sqrt{\hbar/|eB|}$ as unit of length
and $\hbar|eB|/m$ as unit of energy, with $\mu=E\sqrt{m^{2}/|eB^{3}|\hbar}$.
Since $p_{y}$ commutes with $H^{0}$, there is a direct integral decomposition of
$H^{0}=\int^{\oplus}{\rm d} k H_{k}^{0}$, with
\be
H_{k}^{0}=-\frac{1}{2}\frac{{\rm d}^{2}}{{\rm d}x^{2}}+\frac{1}{2}(k-x)^{2}-\mu x,  \,\,\,k\in \reel
\label{2}
\ee
The spectral representation of $H^{0}$ is explicitly given in terms of the 
generalized eigenfunctions
\be
(xy\ket{nk}=\frac{1}{\sqrt{2\pi}}{\rm e}^{iky}u_{n}\(x-k-\mu\),\;\;
n=0,1,\ldots,\;\,\;k\in \reel
\label{3}
\ee
and spectral branches
\be
\epsilon_{n}(k)=\(n+\frac{1}{2}\)-\mu k-\frac{\mu^{2}}{2}
\label{4}
\ee
In (\ref{3}), $u_{n}$ are the usual normalized eigenfunctions of an harmonic oscillator of frequency equal to one
\be
u_{n}(x)=\(\frac{1}{\sqrt{\pi}2^{n}n!}\)^{1/2} {\rm e}^{-\frac{1}{2}x^{2}}H_{n}(x)
\label{4a} 
\ee
with $H_{n}$ the $n^{{\rm th}}$ Hermite polynomial.
For $\mu\neq 0$, the branches $\epsilon_{n}(k)$ 
are linear functions of $k$, thus without points of constancy, implying that the spectrum of $H^{0}$
is absolutely continuous on $\reel$ [3]. If $\mu=0$, $\epsilon_{n}(k)=n+\frac{1}{2}$ are constant,
the spectrum reduces to the infinitely degenerated Landau levels and $k$ labels the degeneracy
in the corresponding subspaces.

The total Hamiltonian $H$ is obtained by formally adding to $H^{0}$ the singular potential $V(x,y)=
\lambda\delta(x,y)$, where $\delta(x,y)$ is the two dimensional Dirac function.
It is well known that this singularity is too strong in $2D$ and a 
renormalization of the coupling constant is needed [4, 5].
Introducing the resolvents $R_{z}=(H-z)^{-1}$ and $R^{0}_{z}=(H^{0}-z)^{-1}$, $z\in{\cal C}$, and
solving the resolvent equation $R_{z}=R^{0}_{z}-R^{0}_{z}VR_{z}$
leads to
\begin{eqnarray}
R_{z}&=&R_{z}^{0}-\frac{R^{0}_{z}|00)(00|R^{0}_{z}}{g(z)}\label{5}\\
g(z)&=&\lambda^{-1}+(00|R^{0}_{z}|00)
\label{5a}
\end{eqnarray}
and from the spectral decomposition (\ref{3}) of $H^{0}$
\be
(00|R^{0}_{z}|00)=\sum_{n=0}^{\infty}\int {\rm d} k\frac{|(00\ket{nk}|^{2}}
{\epsilon_{n}(k)-z}
\label{6}
\ee
The $n$-summation in (\ref{6}) diverges logarithmically since $\epsilon_{n}(k)\sim n,\;n\to \infty$.
Following the same procedure as in [5], it can be made finite by substracting out the diverging part
and combining it with the coupling constant (using also the normalization of the functions (\ref{3}))
\begin{eqnarray}
g(z)&=&\lim_{N\to\infty}\(\lambda^{-1}+\frac{1}{2\pi}\sum_{n=0}^{N}\frac{1}{n+\frac{1}{2}}\)\nonumber\\
&+&\lim_{N\to \infty} 
\sum_{n=0}^{N}\int {\rm d} k|(00\ket{nk}|^{2}\(\frac{1}{\epsilon_{n}(k)-z}-\frac{1}{n+\frac{1}{2}}\)
\label{7}
\end{eqnarray}
In the first term of (\ref{7}) one chooses $\lambda=\lambda_{N}<0$ be $N$-dependent and negative, and requires that $\lambda_{N}\to
0$ in such a way that $\lim_{N\to\infty}\(\lambda^{-1}_{N}+\frac{1}{2\pi}\sum_{n=0}^{N}\frac{1}{n+\frac{1}{2}}\)=\lambda_{r}$ where
$\lambda_{r}$ is a finite renormalized coupling constant. The limit of the second
term in (\ref{7}) exists (see appendix A) so that the model is defined by its resolvent (\ref{5}) setting
\be
g(z)=\lambda_{r}+
\sum_{n=0}^{\infty}\int {\rm d} k|(00\ket{nk}|^{2}\(\frac{1}{\epsilon_{n}(k)-z}-\frac{1}{n+\frac{1}{2}}\)
\label{8}
\ee
Its spectrum will be determined by the nature of the singularities of $R_{z}$ as $z$ approaches the real axis.
When $\mu=0$, the model reduces to that studied in [5]:
\be
g(z)|_{\mu=0}=\lambda_{r}+\frac{1}{2\pi}\sum_{n=0}^{\infty}\(
\frac{1}{(n+\frac{1}{2})-z}-\frac{1}{n+\frac{1}{2}}\)
\label{8a}
\ee
has one zero $\epsilon_{j}$ in-between each of the Landau levels, 
$j+\frac{1}{2}<\epsilon_{j}< j+\frac{3}{2},\; j=0,1,\ldots$, and a zero $\epsilon_{g}<\frac{1}{2}$ lying below all Landau levels.
These zeros give poles in $R_{z}|_{\mu=0}$ that correspond to nondegenerate eigenvalues $\epsilon_{j}$ of $H|_{\mu=0}$
with normalized eigenvectors $\psi_{j}$ (impurity states). As $z\to \epsilon_{j}$, $g(z)|_{\mu=0}\sim a_{j}(z-\epsilon_{j})$ 
and $(z-\epsilon_{j})\bra{\psi_{j}}R_{z}|_{\mu=0}\ket{\psi_{j}}\to 1$. From (\ref{5}) this determines
the coefficient $a_{j}$ to be  
\be
a_{j}^{-1}=|(00|R^{0}_{\epsilon_{j}}|_{\mu=0}\ket{\psi_{j}}|^{2} 
\label{9}
\ee
The poles coming from $R^{0}_{z}|_{\mu=0}$ correspond to the Landau levels that remain unaffected by the presence of the
impurity (stability of the essential spectrum).

\section{Weak electric field asymptotics}

In presence of the electric field, the boundary values of $g(z)$ as $z=\zeta\pm i\eta$ approaches the real axis are obtained
by an application of the Cauchy principal value formula (writing now explicitely the $\mu$-dependence in the
arguments of the functions)
\be
\lim_{\eta\to 0^{+}}g(\zeta\pm i\eta,\mu)=\alpha (\zeta,\mu)\pm i\beta (\zeta,\mu)
\label{L1}
\ee
with
\be
\alpha (\zeta,\mu)=\lambda_{r}+
\sum_{n=0}^{\infty}{\cal P}\!\!\!\!\!\!\int {\rm d} k|(00\ket{nk}|^{2}\(\frac{1}{\epsilon_{n}(k)-\zeta}-\frac{1}{n+\frac{1}{2}}\)
\label{L2}
\ee
\be
\beta (\zeta,\mu)=\frac{\pi}{\mu}\sum_{n=0}^{\infty}|(00\ket{n\;k_{n}(\zeta,\mu)}|^{2}
\label{L3}
\ee
where $k_{n}(\zeta,\mu)$ solves $\epsilon_{n}(k)=\zeta$, i.e.
\be
k_{n}(\zeta,\mu)=\frac{1}{\mu}\(n+\frac{1}{2}-\zeta-\frac{\mu^{2}}{2}\)
\label{L3a}
\ee
The functions $\alpha (\zeta,\mu)$ and $\beta (\zeta,\mu)$ will determine respectively the spectral shifts and
the lifetimes of the impurity states as $\mu\to 0$. From now on we focus attention on the $j^{{\rm th}}$ impurity
state by studying these functions in a neighborhood  
of the unperturbed energy $\epsilon_{j}$ that does not contain the nearest Landau levels. At $\mu=0$, $\alpha(\zeta,0)$
reduces to the expression (\ref{8a}) and vanishes at $\epsilon_{j}$ with $\frac{\partial\alpha}{\partial
\zeta}(\zeta,0)|_{\zeta=\epsilon_{j}}>0$: the implicit function theorem ensures then that for  $\mu$ small, $\alpha(\zeta,\mu)$ has
a nearby zero at $\zeta =\epsilon_{j}(\mu)$ and
\be
\alpha(\zeta,\mu)=a_{j}(\mu)(\zeta-\epsilon_{j}(\mu))+{\rm O}((\zeta-\epsilon_{j}(\mu))^{2}),\;\;a(\mu)>0
\label{L4}
\ee
The spectral shift $\epsilon_{j}(\mu)-\epsilon_{j}=\epsilon_{j}^{(1)}\mu +{\rm O}(\mu^{2})$ can itself be expanded
in $\mu$ \footnote{One can verify that the linear correction $\epsilon_{j}^{(1)}\mu$ is also obtained by formally applying
the regular perturbation theory to the eigenvalue $\epsilon_{j}$ when the electric field is switched on.} and
$a_j(\mu)= a_{j}+ {\rm O(\mu)}$ where $a_{j}$ has the value (\ref{9}).

We study now the asymptotic behaviour of $\beta (\zeta,\mu)$ as $\mu\to 0$ with $\zeta$ in a neighborhood of $\epsilon_{j}$.
We denote $\Delta_{j}= \min (\epsilon_{j}-(j+\frac{1}{2}),j+\frac{3}{2}-\epsilon_{j})<\frac{1}{2}$ the gap between
$\epsilon_{j}$ and the nearest Landau level: $\Delta_{j}$ can be equal to either quantities depending on the value of
$\lambda_{r}$.  For sake of definitness assume in the sequel that
$\Delta_{j}=\epsilon_{j}-(j+\frac{1}{2})$.

\vspace{3mm}
\noindent{\bf Proposition}

\vspace{2mm}
\noindent Set $\zeta-\frac{\mu^{2}}{2}=j+\frac{1}{2}+\delta,\;\;0<\delta<\frac{1}{2}$.
Then
\be
\beta (\zeta,\mu)=\frac{1}{2\sqrt{\pi}}\frac{2^{j}}{j!}\frac{1}{\mu}\(\frac{\delta}{\mu}\)^{2j}
{\rm e}^{-\(\frac{\delta}{\mu}\)^{2}}  (1+{\rm O}(\mu^{2}))
\label{L5}
\ee

\vspace{3mm}
\noindent{\bf Proof}

 From (\ref{3}), (\ref{4a}), (\ref{L3}) and (\ref{L3a}) $\beta (\zeta,\mu)$ reads  
\be
\beta (\zeta,\mu)=\frac{1}{2\sqrt{\pi}}\frac{1}{\mu}\sum_{n=0}^{\infty}\frac{1}{2^{n}n!} {\rm e}^{-\(\frac{n-j-\delta}{\mu}\)^{2}}
H_{n}^{2}\(\frac{n-j-\delta}{\mu}\)
\label{L6}
\ee
The control of these series necessitates an estimate of the Hermite polynomials when their argument
is of the same magnitude as their order. This is provided by the next lemma (proof in appendix B).

\vspace{2mm}
\noindent{\bf Lemma}
 
Let $x$ be any real number not equal to a positive integer. Then there exists $\mu_{0}>0$ such that
\be
H_{n}\(\frac{n-x}{\mu}\)=2^{n}\(\frac{n-x}{\mu}\)^{n}(1-r_{n}(\mu)),\;\;\,0< \mu \leq \mu_{0}
\label{L7}
\ee
with $r_{n}(\mu)\geq 0$ and $r_{n}(\mu)={\rm O(\mu^{2})}$ as $\mu\to 0$.

\vspace{3mm}

\noindent The lemma gives the upper bound
\be
\beta (\zeta,\mu)\leq \sum_{n=0}^{\infty}b_{n}(\mu)=b_{j}(\mu)\(1+\sum_{n\neq j}^\infty\frac{b_{n}(\mu)}{b_{j}(\mu)}\)
\label{L8}
\ee
with
\be
b_{n}(\mu)=\frac{1}{2\sqrt{\pi}}
\frac{2^{n}}{n!}\frac{1}{\mu}\(\frac{n-j-\delta}{\mu}\)^{2n} {\rm e}^{-\(\frac{n-j-\delta}{\mu}\)^{2}} 
\label{L9}
\ee
For $n\neq j$, the largest Gaussian factor in the ratios $\frac{b_{n}(\mu)}{b_{j}(\mu)}$ occurs 
when $n=j+1$. Factorizing it out, we write these ratios in the form 
\be
\frac{b_{n}(\mu)}{b_{j}(\mu)}=\frac{1}{\mu^{2}} {\rm e}^{-\frac{(1-\delta)^{2}-\delta^{2}}{\mu^{2}}}c_{n}(\mu)
\label{L10}
\ee
One checks from (\ref{L9}) and the above definition  that the  $c_{n}(\mu)$ are bounded as $\mu\to 0$
and $c_{n}(\mu)\leq c_{n}(1)\sim n^{n}e^{-(n^{2}+{\rm O}(n))} $ for $n$ large. Hence 
the series $\sum_{n\neq j}^{\infty}c_{n}(\mu)$ converges and is uniformly bounded with respect to $\mu$, implying with
(\ref{L8}) and (\ref{L10}) that
\be
\beta (\zeta,\mu)\leq b_{j}(\mu)\(1+{\rm O}\({\rm e}^{-\frac{C}{\mu^{2}}}\)\)
\label{L11}
\ee
for some $C>0$. On the other hand, one concludes from (\ref{L6}) and the lemma that
\be
\beta (\zeta,\mu)\geq b_{j}(\mu)(1-r_{j}(\mu))^{2},\;\;\;r_{j}(\mu)={\rm O(\mu^{2})}
\label{L12}
\ee
Combining (\ref{L11}) and (\ref{L12}) gives the result of the proposition. 

\section{Shape of resonances and lifetimes}

The time dependent decay amplitude $\bra{\psi_{j}}\exp(-iHt)\ket{\psi_{j}}$ of the $j^{{\rm th}}$ impurity state 
under a weak electric field is given by the Fourier transform of the density of states
\be
\rho_{j}(\zeta,\mu)=\frac{1}{2i\pi}\lim_{\eta\to 0^{+}}\bra{\psi_{j}}(R_{\zeta+i\eta}-R_{\zeta-i\eta})\ket{\psi_{j}}
\label{S1}
\ee
as $\mu\to 0$.
One finds from (\ref{5})
\begin{eqnarray}
\rho_{j}&=&\frac{1}{2i\pi}\(\frac{f_{+}f^{ *}_{-}}{\alpha-i\beta}
-\frac{f^{ *}_{+}f_{-}}{\alpha+i\beta}\)+\rho_{j}^{0}\nonumber\\
&=&\frac{\beta}{\pi(\alpha^{2}+\beta^{2})}\Re(f_{+}f^{*}_{-})+
\frac{\alpha}{\pi(\alpha^{2}+\beta^{2})}\Im(f_{+}f^{*}_{-})+\rho^{0}_{j}
\label{S2}
\end{eqnarray}
where all the functions depend on the energy $\zeta$; $\rho_{j}^{0}(\zeta,\mu)$ is the corresponding density of states
of the crossed fields Hamiltonian $H^{0}$, $\alpha (\zeta,\mu)$ and $\beta (\zeta,\mu)$ are the functions previously discussed
and
\be
f_{\pm}(\zeta)=\lim_{\eta\to 0^{+}}({00}|R^{0}_{\zeta\pm i\eta}\ket{\psi_{j}}
\label{S3}
\ee
In view of (\ref{L4}) and the fact that $\beta(\zeta,\mu)$ tends to zero, the first term in the right hand 
side of (\ref{S2}) behaves as a Lorentzian in a neighborhood of $\epsilon_{j}$ for $\mu$ small 
\be
\frac{\beta_{j}(\zeta,\mu)}{\pi[(a_{j}(\mu)(\zeta-\epsilon_{j}(\mu)))^{2}+\beta_{j}(\zeta,\mu)^{2}]}\;\Re
(f_{+}(\zeta)f^{*}_{-}(\zeta)) \sim \frac{\frac{1}{2}\Gamma(\mu)}{\pi[(\zeta-\epsilon_{j})^{2}+(\frac{1}{2}\Gamma(\mu))^{2}]}
\label{S4}
\ee
with
\be
\Gamma_{j}(\mu)=2\frac{\beta_{j}(\epsilon_{j},\mu)}{a_{j}}
\label{S5}
\ee
In (\ref{S4}), we have kept the dominant behaviour as $\mu\to 0$ by evaluating $a_{j}(\mu)$ and $\epsilon_{j}(\mu)$
at $\mu =0$ and $\beta(\zeta,\mu),\;f_{\pm}(\zeta)$ at $\zeta =\epsilon_{j}$. The Lorentzian is properly normalized because
$\lim_{\mu\to 0}\Re(f_{+}(\epsilon_{j})f^{*}_{-}(\epsilon_{j}))=
|(00|R^{0}_{\epsilon_{j}}|_{\mu=0}\ket{\psi_{j}}|^{2}=a_{j}^{-1}$ by
(\ref{9}). The last two terms in (\ref{S2}) remain bounded for $\zeta$ in a neighborhood of $\epsilon_{j}$. Moreover,
these two terms vanish as $\mu\to 0$ since $\lim_{\mu\to 0}f_{+}=\lim_{\mu\to 0}f_{-}$ is a real quantity and  $\lim_{\mu\to
0}\rho^{0}_{j}(\zeta,\mu)=0$ when $\zeta$ is in-between  Landau levels.   

In (\ref{S4}), $\Gamma(\mu)=(\tau(\mu))^{-1}$ is the inverse life time of the $j^{{\rm th}}$ resonance
so that by (\ref{S5}) and (\ref{L5}) $\tau_{j}(\mu)$ has the form
\be
\tau_{j}(\mu)=C_{j}\mu\(\frac{\mu}{\Delta_{j}}\)^{2j}\exp\(\frac{\Delta_{j}}{\mu}\)^{2}
\label{S6}
\ee
where $\Delta_{j}=\epsilon_{j}-(j+\frac{1}{2})$. The analysis and the results are similar if
$\Delta_{j}=(j+\frac{3}{2})-\epsilon_{j}$. The lifetime of the resonance corresponding to the lowest energy state $\epsilon_{g}$
is found to be $\tau_{g}(\mu)=C_{g}\mu \exp\(\frac{1/2-\epsilon_{g}}{\mu}\)^{2}$. 

\section{Concluding remarks}

We have shown that all impuritity states are delocalized under the influence of an electric field, 
how weak it may be, but with Gaussian long lifetimes. This has to be compared with pure Stark resonances
that have exponentially long lifetimes [6].

The calculation has been performed with an attractive $\delta$-potential
\footnote{The renormalization procedure given in section
2 requires a negative bare coupling constant.}. We conjecture that it gives the correct indication for the general case:
impurity states due to short range potentials, attractive as well as repulsive,
should have finite (Gaussian long) lifetimes in
presence of a weak electric field. In the repulsive case the only difference is that all impurity states will have energies above
the lowest Landau level. 

To conclude we remark that our results are sensitive to the potential nature of the impurity interaction.
Consider for instance the model $H=H^{0}+\lambda\ket{\phi}\bra{\phi}$ obtained by adding a rank-one interaction to the crossed
fields Hamiltonian, with $\phi(x,y)$ a square integrable function on $\reel^{2}$, namely the impurity interaction 
is represented by the non local separable kernel $\lambda \phi(x,y)\phi^{*}(x^{\prime},y^{\prime})$.
Energies and lifetimes of resonances are now found from the function\footnote{Here renormalization of the coupling constant is
not needed.} 
\be
g_{\phi}(z)=\lambda^{-1}+\bra{\phi}R^{0}_{z}\ket{\phi}
\label{C1}
\ee
in the place of (\ref{5a}) and 
\begin{eqnarray}
\beta_{\phi} (\zeta,\mu)&=&\frac{\pi}{\mu}\sum_{n=0}^{\infty}|\langle \phi\ket{n\;k_{n}(\zeta,\mu)}|^{2}\label{C2}\\
\bra{nk}\phi\rangle&=&\sum_{n=0}^{\infty}\int dx u_{n}(x)\tilde{\phi}(x,k)\label{C3}
\end{eqnarray}
where $\tilde{\phi}(x,k)$ is the Fourier transform of $\phi(x,y)$ with respect to the $y$-coordinate. Suppose now that
\be
\tilde{\phi}(x,k)=0,\;\;\;k\geq k_{0}>0
\label{C4}
\ee
Then when $\mu$ is small enough (say $\mu\ll\frac{\Delta_{j}}{k_{0}}$), there exists a  neighborood of $\epsilon_{j}$
such that $k_{n}(\zeta,\mu)\geq k_{0}$ for all $n$ and thus $\beta_{\phi} (\zeta,\mu)$ vanishes when $\zeta$ is in this
neighborhood. This implies that the $j^{{\rm th}}$ impurity state remains an eigenvector of $H$ with an eigenvalue close to
$\epsilon_{j}$ embedded in the continuum. One sees on this example that the interaction has to be sufficiently non local
since by the condition (\ref{C4}) the support of $\phi(x,y)$ must extend on the whole $y$-axis.  

\vspace{5mm}
\appendix

\section*{Appendix A: Existence of the renormalized model}

\vspace{3mm}
\noindent Introducing (\ref{3}) in (\ref{8}) (using also parity of the function $u_{n}$), we split the sum in two terms
\be
         \sum\limits_{n=0}^{\infty}\int {\rm d} k\left\vert u_{n}(-k)\right\vert^{2}\left(\frac{1}{\epsilon_{n}(k-\mu)-z}-
\frac{1}{n+\frac{1}{2}}\right) = I_{1}+I_{2}
\label{A1}
\ee

\begin{eqnarray*}
        I_{1} &=& \sum\limits_{n=0}^{\infty}\frac{1}{(n+\frac{1}{2})n^{\varsigma}}\int_{\left\vert k \right\vert 
\leq n^{\gamma}} {\rm d} k \left\vert
u_{n}(k)\right\vert^{2}\,n^{\varsigma}\left(\frac{k\mu-\frac{\mu^{2}}{2}+z}{\epsilon_{n}(k-\mu)-z}\right) \\
        I_{2} &=& \sum\limits_{n=0}^{\infty} \int_{\left\vert k \right\vert \geq n^{\gamma}} {\rm d} k\, 
\frac{\left\vert u_{n}(k)\right\vert^{2}}{\epsilon_{n}(k-\mu)-z} - \sum\limits_{n=0}^{\infty} \int_{\left\vert k \right\vert \geq
n^{\gamma}} {\rm d} k\, \frac{\left\vert u_{n}(k)\right\vert^{2}}{n+\frac{1}{2}}  \end{eqnarray*}
with
\be
0<\varsigma<\frac{1}{4},\;\;\;\frac{1}{2}<\gamma=1-2\varsigma<1
\label{A2}
\ee
One sees in the integrand of $I_{1}$ that the fraction
$\left\vert n^{\varsigma}\left(\frac{z+k\mu-\mu^{2}/2}{-z+(n+1/2)-\mu k+\mu^{2}/2}\right) \right\vert < C
<\infty$ is bounded uniformly with respect to $k$ and $n$ for $\left\vert k \right\vert \leq n^{\gamma}$, $\Im z \neq 0$.
Hence, since the $u_{n}$ are normalized, $I_{1}\leq 
C\sum\limits_{n=0}^{\infty}\frac{1}{(n+\frac{1}{2})n^{\varsigma}}<\infty$.
For $I_{2}$, since $\frac{1}{\left\vert \epsilon_{n}(k-\mu)-z \right\vert} \leq
\frac{1}{\left\vert \Im z \right\vert}$, it is sufficient to show that $\sum\limits_{n=0}^{\infty} \int_{\left\vert k
\right\vert \geq n^{\gamma}} {\rm d} k \left\vert u_{n}(k)\right\vert^{2} < \infty$.
From the integral representation of the Hermite polynomials [6]
\be
H_{n}(y) = 2^{n}\frac{1}{\sqrt{\pi}}\,\int\limits_{-\infty}^{+\infty} {\rm d} t\,(y+it)^{n}\,
{\rm e}^{-t^2}
\label{A3}
\ee
one deduces (using $\log(1+y) \leq y$)  
\be
        H_{n}(k) \leq \frac{2^{n}k^{n}}{\sqrt{\pi}} \int {\rm d} t\, \left(1+\left\vert \frac{t}{k} \right\vert \right)^{n}\,
{\rm e}^{-t^{2}} \leq \frac{2^{n}k^{n}}{\sqrt{\pi}} \int {\rm d} t\, {\rm e}^{n \left\vert \frac{t}{k} \right\vert}\,
{\rm e}^{-t^{2}} \leq 2^{n+1}k^{n}\, {\rm e}^{\frac{n^{2}}{4k^{2}}}
\label{A4}
\ee
Noting that $n$ sufficiently large ($n\geq n_{0}$) and $\left\vert k \right\vert > n^{\gamma},\gamma>\frac{1}{2}$, 
implies ${\rm e}^{-\frac{k^{2}}{2}
(1-\frac{n^{2}}{2k^{4}})} \leq {\rm e}^{-\frac{k^{2}}{4}}$, one finds from (\ref{A4})
\begin{eqnarray}
        \sum\limits_{n=n_{0}}^{\infty} \int_{\left\vert k \right\vert \geq n^{\gamma}} {\rm d} k \left\vert u_{n}(k)
\right\vert^{2} &=& \frac{1}{2\pi\sqrt{\pi}}\sum\limits_{n=n_{0}}^{\infty}\frac{1}{2^{n}n!} \int_{\left\vert k \right\vert \geq
n^{\gamma}} {\rm d} k\, {\rm e}^{-k^{2}}\,H_{n}^{2}(k)\nonumber \\
        &\leq& \frac{1}{2\pi\sqrt{\pi}}\sum\limits_{n=n_{0}}^{\infty} \frac{2^{n}}{n!}\,{\rm e}^{-\frac{n^{2\gamma}}{2}} 
\int_{\left\vert k \right\vert \geq n^{\gamma}} {\rm d} k\, {\rm e}^{-\frac{k^{2}}{4}}\,k^{2n} \nonumber\\
        &\leq& \frac{8{\rm e}^{4}}{\pi}\sum\limits_{n=n_{0}}^{\infty} 2^{n}n!\;
{\rm e}^{-\frac{n^{2\gamma}}{2}} < \infty 
\label{A5}
\end{eqnarray}
where the last inequality follows from $|k|^{n}\leq n!\,{\rm e}^{|k|} $. The series (\ref{A5}) converges for
$\gamma>\frac{1}{2}$. 

\vspace{5mm}
\section*{Appendice B: Proof of the lemma}

\vspace{3mm}
\noindent The lemma is true by inspection for the cases $n=1,2,3$. If $n\geq 4$
one uses the formula (\ref{A3}) 
\be
        H_{n}\left(\frac{n-x}{\mu}\right) = 2^{n}\left(\frac{n-x}{\mu}\right)^{n} f_{n}(\mu),\;\;\;
f_{n}(\mu) = \frac{1}{\sqrt{\pi}}\,\int{\rm d} t\,\left(1+\frac{ i\mu t}{n-x}\right)^{n}\,{\rm e}^{-t^2}
\ee
The limited Taylor expansion of $f_{n}(\mu)$ around $\mu=0$ gives 
\be
        f_{n}(\mu) = 1 + \frac{1}{2!}\mu^2\,f_{n}^{''}(0) + \frac{1}{4!}\mu^{4}\,f_{n}^{''''}(\overline{\mu}),\qquad 0
\leq\overline{\mu} \leq \mu 
\ee
with
\begin{eqnarray*}
        f_{n}^{''}(0) &=& -\frac{n(n-1)}{(n-x)^{2}} \\
        f_{n}^{''''}(\overline{\mu}) &=& \frac{n!}{(n-4)!(n-x)^{4}}\,\frac{1}{\sqrt{\pi}}\int{\rm d} t\,t^{4}
\left(1+ i\frac{\overline{\mu} t}{n-x}\right)^{n-4}\,{\rm e}^{-t^{2}}
\end{eqnarray*}
If $x$ is not a positive integer, there exists $C_{1}$ independent of $n$, $0<C_{1}<\infty$ such that
$f_{n}^{''}(0)\geq -C_{1}$. Moreover, one has for $n\geq 4$ and using $\log(1+y) \leq
y$   
\begin{eqnarray*}
        \left\vert f_{n}^{''''}(\overline{\mu}) \right\vert &\leq&  \frac{n!}{(n-4)!(n-x)^{4}}\,\frac{1}{\sqrt{\pi}}
\int{\rm d} t\,t^{4}\left(1+\overline{\mu} \left\vert \frac{t}{n-x} \right\vert\right)^{n-4}\,{\rm e}^{-t^{2}} \\
        &\leq& \frac{n!}{(n-4)!(n-x)^{4}}\,\frac{1}{\sqrt{\pi}}\int{\rm d} t\,t^{4}\,{\rm e}^{\frac{n-4}{\left\vert n-x 
\right\vert}\,\overline{\mu} \left\vert t \right\vert}\,{\rm e}^{-t^{2}} \leq C_{2}<\infty 
\end{eqnarray*}
with $C_{2}$ independent of $n$ and of $\overline{\mu}$ in compact sets. This leads to the conclusion of the lemma.

\end{document}